\documentstyle[aps,prl,multicol,epsf]{revtex}
\begin{document}
\title{Single-particle spectra near a stripe instability}  
\author{ S. Caprara, C. Di Castro, and M. Grilli}  
\address{Istituto Nazionale per la Fisica della Materia, UdR Roma 1 and  
Dipartimento di Fisica, Universit\`a di Roma ``La Sapienza'',\\  
Piazzale Aldo Moro 2, I-00185 Roma, Italy}  
\maketitle

\begin{abstract} 
We analyze the single-particle spectra of a bi-layered electron system
near a stripe instability and compare the results with ARPES experiments on 
the Bi2212 cuprate superconductor near optimum doping, addressing also the 
issue of the puzzling absence of bonding-antibonding splitting.
\end{abstract}  

\begin{multicols}{2}

It was proposed that the anomalous normal-state properties of the 
cuprate superconductors at optimum doping may result from the 
mixing of the doped holes with the collective charge and spin fluctuations 
near a stripe instability \cite{cdg}. To investigate the corresponding
quasiparticle spectra, and to compare with experiments on the bi-layered 
Bi2212 \cite{saini}, we introduce the Hamiltonian
\begin{eqnarray*}
H&=&\sum_{k;\sigma}\sum_\ell\left(\xi_k-\delta\mu\right) 
c^{+}_{k\sigma\ell}c_{k\sigma\ell}\\
&-&t_\perp \sum_{k;\sigma} \gamma_k 
\left( c^+_{k\sigma1}c_{k\sigma2} + h.c.\right)\\
&+&
\sum_{k,q;\sigma,\rho}\sum_\ell \sum_i g_i c^{+}_{k+q\sigma\ell}c_{k\rho\ell}
\tau^i_{\sigma\rho}S^i_{-q\ell}
\end{eqnarray*}
where $\xi_k=-2t(c_x+c_y)+4t'c_x c_y-\mu$, with $c_{x,y}=\cos(k_{x,y}a)$,
is the tight-binding dispersion for electrons on a square 
lattice with nearest and next-to-nearest neighbors hopping, $a$ is the 
lattice spacing, $\mu$ is the bare chemical potential and $\delta\mu$ is the 
shift in the interacting system. The planes are labelled by $\ell=1,2$  
and $t_\perp$ is the interplane matrix element modulated by
$\gamma_k={1\over 2}|c_x-c_y|$ to account for the suppression 
at $k_x\simeq \pm k_y$ \cite{and}. 
The constants $g_i$ couple electrons to charge ($i=0$) and spin ($i=1,2,3$) 
fluctuations. The spin structure of the generalized density coupled to 
$S_{-q\ell}^i$ is accounted for by the Pauli matrix $\tau^i$. The counterterm
$\delta\mu\sim O(g^2)$ is determined, 
order by order in perturbation theory, to fix the number of electrons.
The fluctuating fields $S_{-q\ell}^i$ are characterized  by the 
susceptibilities $\chi_{ij\ell\ell'}(q,\omega)=\delta_{ij} 
\delta_{\ell\ell'} A_i/[\kappa^2+\eta_{q-Q_{i\ell}}-{\rm i}{\bar \tau} 
\omega]^{-1}$,
where $A_i$ are constants, the mass $\kappa^2$ vanishes at 
criticality, $\eta_k=2-c_x-c_y$ reproduces the $k^2$ behavior 
at $ka\ll 1$, preserving the lattice periodicity, $Q_{i\ell}$ are 
the critical wave-vectors, ${\bar \tau}$ is a 
characteristic time scale, and the dimensionless coupling 
constants are $\lambda_i=g_i^2 A_i/t$.
In the paramagnetic phase the parameters are the same for $i=1,2,3$, and we 
label charge and spin fluctuations with $c,s$. The direction of 
the charge modulation is debated. As a matter of illustration we analyze the 
case $Q_{c1}= 0.4(\pi/a,-\pi/a)$, $Q_{s1}=(\pi/a,\pi/a)$, suggested by 
the ARPES experiments on Bi2212 \cite{saini}. We allow for a 
mismatch of the charge modulation pattern on the two planes and take 
$Q_{c2}= 0.4(\pi/a,\pi/a)$, while $Q_{s2}=Q_{s1}$.
The Hamiltonian is written in the simpler ``plane representation'', in which
the interaction is diagonal, and the decoupling of the two planes as 
$t_\perp\to 0$ is evident. The alternative ``band representation'' is 
obtained by diagonalization of the fermionic part of the Hamiltonian, but the 
interaction is not diagonal, and the role of $t_\perp$ is less transparent.
The two representations are, of course, equivalent.

The $O(\lambda)$ perturbation theory accounts for the main dressing of the 
electrons. An average over $\pm Q_c$ is performed to maintain inversion 
symmetry, leading to the self-energy $\Sigma_\ell=<\Sigma_{c\ell}>+3\Sigma_s
-\delta\mu$. Previous analysis for $t_\perp=0$ \cite{pap} showed that 
spectral weight is transferred from the quasiparticle peak to incoherent 
shadow resonances. The changes in the single-particle spectra and in the 
distribution of low-lying spectral weight are in agreement with the 
experiment \cite{saini}. The suppression of spectral weight at the Fermi 
level near the $M$ points [i.e. $(0,\pm\pi/a)$ and $(\pm\pi/a,0)$], due to 
spin fluctuations, is modulated by charge fluctuations. 

Here we address the absence of bi-layer splitting, despite a sizable 
calculated $t_\perp$ \cite{and}. The contribution of the plane $\ell$ to the 
spectral density is
$$
A_\ell =-{1\over \pi}\sum_{\alpha=1,2}
{\rm Im} \left. {1+(-1)^{\alpha+\ell}\Sigma_- /{\tilde t}_\perp
\over \omega - \xi_k-\Sigma_+ +(-1)^\alpha {\tilde t}_\perp}
\right|_{R}
$$
where $\Sigma_\pm={1\over 2}(\Sigma_{\ell=1}\pm\Sigma_{\ell=2})$,
${\tilde t}_\perp=(t_\perp^2\gamma_k^2+\Sigma_-^2)^{1/2}$ and ${R}$ means 
retarded. For $\lambda_i=0$ the quasiparticle FS consists of two branches, 
corresponding to the bonding and antibonding band 
(Fig. 1, left panel), well separated near the M points, 
where $\gamma_k$ is larger. However, a moderate coupling between the 
electrons and the critical fluctuations is sufficient to eliminate
the FS splitting. Indeed, the 
suppression of spectral weight is stronger near the $M$ points where the 
splitting is expected to be larger, and is weaker where the splitting is 
naturally suppressed by $\gamma_k$. The effect is enhanced in the case of a 
mismatch between the fluctuation patterns on the two different planes. The 
resulting FS, projected onto a single plane (e.g. $\ell=1$) as it is suitable 
to interpret ARPES results, is essentially the same as in the absence of 
interlayer coupling (Fig. 1, right panel). Thus the absence of the band 
splitting in ARPES spectra of bi-layered materials can be due to the 
enhancement of charge and spin fluctuations scattering the quasiparticles 
near a stripe instability.

{ Acknowledgments.} 
Part of this work was carried out with the financial 
support of the I.N.F.M. - P.R.A. 1996.

   
\end{multicols}  
\begin{figure}[htbp]
\begin{center}
\setlength{\unitlength}{1truecm}
\begin{picture}(5.0,15.0)
\put(-4.0,-8.0){\epsfbox{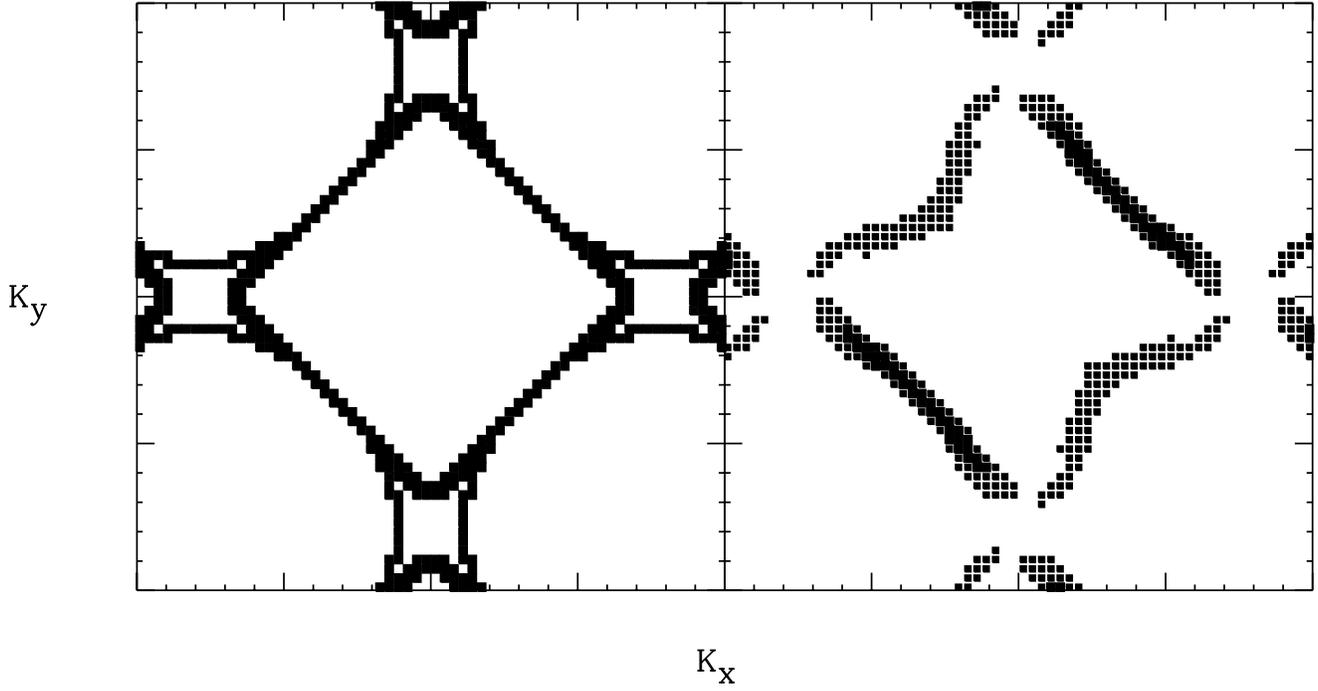}}
\end{picture}
\end{center}
\caption{Calculated FS for $t=200$ meV, $t'=50$ meV, 
$t_\perp=50$ meV, $\mu=-180$meV, ${\bar \tau}^{-1}=200$ meV and 
$\lambda_{s,c}=0$ (left), $\lambda_s=0.5$, $\lambda_c=0.3$ (right).}
\protect\label{fig1}
\end{figure}

\end{document}